\documentclass[aps,prd,onecolumn,groupedaddress,showpacs,nofootinbib,amssymb]{revtex4}
\usepackage{amsmath}
\usepackage{amssymb}
\usepackage{amsfonts}
\usepackage{graphicx,bm}
\usepackage{color,amsxtra}
\usepackage{epsf}
\usepackage{enumerate}
\usepackage{hhline}
\usepackage{array}
\usepackage{tabularx}
\newcommand{\be}{\begin{equation}}
\newcommand{\ee}{\end{equation}}
\newcommand{\bea}{\begin{eqnarray}}
\newcommand{\eea}{\end{eqnarray}}
\newcommand{\beaa}{\begin{eqnarray*}}
\newcommand{\eeaa}{\end{eqnarray*}}

\newcommand{\nn}{\nonumber \\}
\newcommand{\e}{\mathrm{e}}

\def\be{\begin{equation}}
\def\ee{\end{equation}}
\def\bea{\begin{eqnarray}}
\def\eea{\end{eqnarray}}

\def\nn{\nonumber \\}
\def\e{\mathrm{e}}

\begin{document}

\tolerance=5000

\title{Dynamical Domain Wall and Localization}

\author{Yuta Toyozato$^{1, }$\footnote{
E-mail address: toyozato@th.phys.nagoya-u.ac.jp},  Masafumi Higuchi$^{1, }$\footnote{
E-mail address: mhiguchi@th.phys.nagoya-u.ac.jp}, 
Shin'ichi Nojiri$^{1, 2,}$\footnote{E-mail address:
nojiri@phys.nagoya-u.ac.jp}
}

\affiliation{
$^1$ Department of Physics, Nagoya University, Nagoya
464-8602, Japan \\
$^2$ Kobayashi-Maskawa Institute for the Origin of Particles and
the Universe, Nagoya University, Nagoya 464-8602, Japan}

\begin{abstract}

Based on the previous works (arXiv:1202.5375 and 1402.1346), we investigate the localization of the fields on the dynamical domain wall, where the four dimensional FRW universe is realized on the domain wall in the five dimensional space-time. 
Especially we show that the chiral spinor can localize on the domain wall, which has not been succeeded in the past works as the seminal work in arXiv:0810.3746. 

\end{abstract}


\maketitle

\section{Introduction \label{I}}

The scenarios that our universe could be a brane or domain wall embedded in 
a higher dimensional space-time are not new \cite{Akama:1982jy, Rubakov:1983bb} 
but after we have found the strong conjecture that there could be a classical solution 
called $D$-brane in string theories \cite{Dai:1989ua, Polchinski:1996na}, the scenarios of 
the brane 
\cite{Randall:1999ee, Randall:1999vf, Dvali:2000hr, Deffayet:2000uy, Deffayet:2001pu}
or the domain wall 
\cite{Lukas:1998yy, Kaloper:1999sm, Chamblin:1999ya, DeWolfe:1999cp, Gremm:1999pj, Csaki:2000fc,Gremm:2000dj, Kehagias:2000au, Kobayashi:2001jd, Slatyer:2006un, George:2008vu}. 
with reality have been well investigated. 
Especially inflationary brane world models by using the trace anomaly have been proposed 
\cite{Nojiri:2000eb,Hawking:2000bb,Nojiri:2000gb}. 
We may regard brane with a limit where the thickness of the domain wall vanishes. 

Recently some of the authors proposed a domain wall model with two scalar 
fields Refs.~\cite{Toyozato:2012zh, Higuchi:2014bya}, we have shown to be able to construct a 
model which generates space-time, where the scale factor of the domain wall universe, which 
could be the general FRW universe, and the warp factor are arbitrarily given. 
The formulation is an extension of the formalism of the reconstruction of the domain 
wall \cite{Nojiri:2010wj}. 
We should note that in \cite{DeWolfe:1999cp}, it has been proposed a formulation where only 
the warp factor of the domain wall is arbitrary. 

In this paper, we consider the localization of several fields in the model \cite{Toyozato:2012zh, Higuchi:2014bya}. 
The localization of the graviton has been already shown in \cite{Higuchi:2014bya} and we have 
found that there appear extra terms which is related with the extra dimension. This terms may 
affect the fluctuations of the universe. 
We now investigate the localization of the spinor field and the vector field in this paper. 
If the domain is static, there appear chiral fermions localized on the domain wall  \cite{Kehagias:2000au}. 
There were attempts to localize the chiral fermion on the dynamical domain wall corresponding 
to the FRW universe \cite{George:2008vu} but it was not succeeded. 
In \cite{George:2008vu}, it was assumed that the warp factor does not depend on the time but 
in this paper, we use the time-dependent warp factor and show the localization of the chiral 
fermion. 
In this paper, $A$, $B$ denote  indices on a curved space-time manifold, while $\hat{A}, \hat{B}$ denote indices on its tangent space-time.

In the next section, we review on the formulation in \cite{Toyozato:2012zh, Higuchi:2014bya} including the localization of the graviton. 
In section \ref{III}, we investigate the localization of the chiral fermion on the domain wall. 
In section \ref{IV}, the localization and non-normalizability of vector field are shown.
The final section, section \ref{V}, is devoted to the discussions. 

\section{Domain wall model with two scalar fields \label{II}}

We now briefly review on the formulation of the domain wall model with two scalar fields 
based on \cite{Toyozato:2012zh, Higuchi:2014bya} and we show the localization of the graviton. 
The formulation is given by extending the formulation in \cite{Bamba:2011nm} and similar 
procedure was invented for the reconstruction of the FRW universe by single scalar model 
\cite{Capozziello:2005tf}. 

In the domain wall solution, which can be regarded as a general FRW universe in the five dimensional space-time, the metric is given by
\be
\label{metricR}
ds^2 = dw^2 + L^2 \e^{u \left( w,t \right)} ds_\mathrm{FRW}^2 \, . 
\ee
Here $ds_\mathrm{FRW}^2$ is the metric of the general FRW universe, 
\be
\label{FRWmetric0}
ds_\mathrm{FRW}^2 = - dt^2 + a \left( t \right)^2 \left\{   \frac{dr^2}{1-kr^2} 
+ r^2 d \theta^2 +r^2 \sin^2 \theta d \phi^2  \right\}\, .
\ee
The FRW space-time is embedded by the arbitrary warp factor $L^2 \e^{u \left( w,t \right)}$. 

In \cite{Toyozato:2012zh, Higuchi:2014bya}, we considered the following action with two scalar fields $\phi$ and $\chi$: 
\be
\label{pc1}
S_{\phi\chi} = \int d^5 x \sqrt{-g} \left\{ \frac{R}{2\kappa^2} 
 - \frac{1}{2} A (\phi,\chi) \partial_M \phi \partial^M \phi 
 - B (\phi,\chi) \partial_M \phi \partial^M \chi 
 - \frac{1}{2} C (\phi,\chi) \partial_M \chi \partial^M \chi - V (\phi,\chi)\right\}\, .
\ee
It has been shown that we can construct a model to realize the arbitrary metric 
(\ref{metricR}) by using the model (\ref{pc1}).
In the model (\ref{pc1}), the energy-momentum tensor for the scalar fields $\phi$ and $\chi$ are given by
\begin{align}
\label{pc2}
T^{\phi\chi}_{MN} =& g_{MN} \left\{ 
 - \frac{1}{2} A (\phi,\chi) \partial_L \phi \partial^L \phi 
 - B (\phi,\chi) \partial_L \phi \partial^L \chi 
 - \frac{1}{2} C (\phi,\chi) \partial_L \chi \partial^L \chi - V (\phi,\chi)\right\} \nn
& +  A (\phi,\chi) \partial_M \phi \partial_N \phi 
+ B (\phi,\chi) \left( \partial_M \phi \partial_N \chi 
+ \partial_N \phi \partial_M \chi \right) 
+ C (\phi,\chi) \partial_M \chi \partial_N \chi \, .
\end{align}
On the other hand, by the variation of $\phi$ and $\chi$, we obtain the field equations as follows, 
\begin{align}
\label{pc3}
0 =& \frac{1}{2} A_\phi \partial_M \phi \partial^M \phi 
+ A \nabla^M \partial_M \phi + A_\chi \partial_M \phi \partial^M \chi 
+ \left( B_\chi - \frac{1}{2} C_\phi \right)\partial_M \chi \partial^M \chi  
+ B \nabla^M \partial_M \chi - V_\phi \, ,\\
\label{pc4}
0 =& \left( - \frac{1}{2} A_\chi + B_\phi \right) \partial_M \phi \partial^M \phi 
+ B \nabla^M \partial_M \phi 
+ \frac{1}{2} C_\chi \partial_M \chi \partial^M \chi 
+ C \nabla^M \partial_M \chi + C_\phi \partial_M \phi \partial^M \chi 
 - V_\chi\, .
\end{align}
Here we use the notation $A_\phi=\partial A(\phi,\chi)/\partial \phi$, etc. 
We now choose $\phi=t$ and $\chi=w$. Then we find
\be
\label{pc4b}
T_0^{\ 0} = - \frac{\e^{-2u \left( w,t \right)} }{2L^2 } A - \frac{1}{2} C - V\, ,\quad 
T_i^{\ j} = \delta_i^{\ j} \left( \frac{\e^{-2u \left( w,t \right)} }{2L^2 } A - \frac{1}{2} C - V \right)\, ,\quad 
T_5^{\ 5} = \frac{\e^{-2u \left( w,t \right)} }{2L^2 } A + \frac{1}{2} C - V\, ,\quad 
T_0^{\ 5} = B \, ,
\ee
By using the Einstein equation, we may solve Eqs.~(\ref{pc4b}) with respect to $A$, $B$, $C$, and $V$ as follows,  
%
% by choosing 
\begin{align}
\label{pc7}
A =& \frac{L^2 \e^{u \left( w,t \right)}}{\kappa^2 } \left( G_1^{\ 1} - G_0^{\ 0} \right) 
=  \frac{L^2 \e^{u \left( w,t \right)}}{\kappa^2 } \left( G_2^{\ 2} - G_0^{\ 0} \right) 
=  \frac{L^2 \e^{u \left( w,t \right)}}{\kappa^2 } \left( G_3^{\ 3} - G_0^{\ 0} \right) \nn
=& \frac{1}{\kappa^2} \left( \frac{2k}{a^2} - \ddot u - 2 \dot H + \frac{\left( \dot u \right)^2}{2} + \dot u H \right) \, , \nn
B = & \frac{1}{\kappa^2}G_0^{\ 5} = - \frac{3u'}{2 \kappa^2 L^2 \e^u} 
\left( \dot u + 2 H \right) \, , \nn
C =& \frac{1}{\kappa^2} \left( G_5^{\ 5} - G_1^{\ 1} \right)
= \frac{1}{\kappa^2} \left( G_5^{\ 5} - G_2^{\ 2} \right) 
= \frac{1}{\kappa^2} \left( G_5^{\ 5} - G_3^{\ 3} \right) \nn
=& \frac{1}{\kappa^2} \left( - \frac{3}{2} u'' + \frac{2k}{L^2 \e^u a^2} 
- \frac{1}{2\e^u} \left( \ddot u + 2 \dot H + \left( \dot u \right)^2 +5\dot u H + 6 H^2 \right) \right)\, , \nn
V = & \frac{1}{\kappa^2} \left( G_0^{\ 0} + G_5^{\ 5} \right) \nn
=& \frac{1}{\kappa^2} \left( - \frac{3}{4} \left( u'' + 2 \left( u' \right)^2 \right) 
+ \frac{3k}{L^2 \e^u a^2 } + \frac{1}{4L^2 \e^u} \left( 3 \ddot u + 6 \dot H + 3 \left( \dot u \right)^2 + 15 \dot u + 18 H^2 \right) \right)\, .
\end{align}
Here the Einstein tensor is denoted by $G_{MN}$. 
%In the present model, the solution for the scalar field is given by $\phi=t$ and $\chi=w$. 
Then by replacing $t$ and $w$ in the r.h.s. of Eqs.~(\ref{pc7}) by $\phi$ and $\chi$, respectively, we find the explicit forms of $A(\phi,\chi)$, $B(\phi,\chi)$, $C(\phi,\chi)$, and $V(\phi,\chi)$. 
Then by using the expressions in the action (\ref{pc1}), we obtain a model which realize the metric (\ref{metricR}). 
%%%%%%%%%%
We should note that Eqs.~(\ref{pc3}) and (\ref{pc4}) are satisfied automatically because they can be obtained by using the Einstein equation and the 0th and 5th components of the Bianchi identity $\nabla^N \left( R_{MN} - \frac{1}{2} R g_{MN} \right) = 0$, which corresponds to the conservation of the energy-momentum tensors in (\ref{pc4b}). 
%%%%%%%%%%

The localization of the graviton can be investigated by considering the perturbation
\be
\label{H1}
g_{MN} \to g_{MN} + h_{MN}\, .
\ee
We now impose the gauge condition
\be
\label{H4}
\nabla^M h_{MN} = g^{MN} h_{MN} = 0\, .
\ee
In our domain wall model, by writing $h_{ij}$ ($i,j=1,2,3$) in a factorized form, $h_{ij}(w,x)= \e^{u(w,t)}\hat{h}_{ij}(x)$, we find that the graviton follows the equation,
\be
\label{LB6}
0 = \left(2\frac{\dot{a}\dot{u}}{a}-\dot{u}\partial_0+2\frac{\ddot{a}}{a}
+\frac{\dot{a}}{a}\partial_0-\partial_0^2+\frac{\bigtriangleup}{a^2}\right)\hat{h}_{ij}\, .
\ee
Here $\bigtriangleup$ is the Laplacian in the flat three dimensional space. 
Then if $u$ goes to minus infinity rapidly enough for large $|w|$, $h_{ij}(w,x)$ can be 
normalized in the direction of $w$ and therefore the graviton localizes on the domain wall. 

We should note that the graviton $h^{(4)}_{ij}$ in the four dimensional FRW space-time (\ref{FRWmetric0}), satisfies the following equation  
\be
\label{LB3}
0=\left(2\frac{\ddot{a}}{a}+\frac{\dot{a}}{a}\partial_0-\partial_0^2 
+\frac{\bigtriangleup}{a^2}\right)h^{(4)}_{ij}\, ,
\ee
in the standard four dimensional Einstein gravity. 
Therefore when $\dot{u}\left(2\frac{\dot{a}}{a}-\partial_0\right)\hat{h}_{ij}=0$, 
the expression (\ref{LB6}) coincides with the equation for the graviton in (\ref{LB3}). 
Especially when the warp factor does not depend on time-that is, $\dot u=0$, the two 
expressions coincide with each other. 
Conversely if $\dot u \neq 0$, corrections proportional to $\dot u$ are possible when we consider the perturbations. 

We have only considered the perturbation of the graviton, which is a tensor field and therefore the graviton does not mix with the scalar modes. This is because we are interested in the localization of the graviton. Of course, it is necessary to consider the perturbations including the scalar modes when we consider more realistic cosmology but the perturbations become complicated and we like to leave the perturbations including the scalar modes in the future works.     

\section{Localization of Fermion \label{III}}

In this section, we show that chiral fermion can localize on the domain wall. 
The Dirac equation in five dimensions is given by
\begin{equation}
\label{YT1}
\Gamma^{M} \nabla_{M} \Psi + \tilde{f} \chi \left( w \right) \Psi = 0\, .
\end{equation}
In the last term, $\tilde{f} \chi$ is a function of the scalar field $\chi=w$ in (\ref{pc1}) and this 
term expresses the general yukawa interaction between the fermion and the scalar field $\chi$. 
In (\ref{YT1}), $\nabla_M$ is a covariant derivative: 
$\nabla_{M} \overset{\mathrm{def}}{=} \partial_{M} + \frac{1}{4} \omega_{ABM} \Gamma^{AB}$, 
$\Gamma^{AB} \overset{\mathrm{def}}{=} \frac{1}{2} \left[   \Gamma^A, \Gamma^B \right]$. 
By decomposing $\Psi$ as $\Psi = \eta \left( t, w \right) \psi \left( x \right)$, 
where $\psi$ corresponds to the 4-dimensional Dirac spinor, we find 
\begin{align}
\label{YT2}
\Gamma^{M} \nabla_{M} \Psi + \tilde{f} \chi \left( w \right) \Psi =&  \left( \Gamma^{\mu} 
\partial_{\mu} \psi  \right) \eta + \Gamma^5 \psi \left( \partial_5 \eta \right) 
+ \Gamma^0 \psi \left( \partial_0 \eta \right) + \frac{1}{4} \Gamma^M \omega_{ABM} 
\Gamma^{AB} \Psi + \tilde{f} \chi \Psi \nonumber \\
=&  L^{-1}  \e^{-u/2} \left\{  \Gamma^{\hat{0}}  \partial_0 \psi + \frac{1}{ a \left( t \right)} 
\Gamma^{\hat{i}} \partial_i \psi \right\} \eta + \Gamma^{\hat{5}} \psi  \left( \partial_5 \eta 
\right)  + L^{-1} \e^{-u/2} \Gamma^{\hat{0}} \psi \left( \partial_0 \eta \right) \nonumber \\
& +  u^{\prime} \Gamma^{\hat{5}} \psi \eta + \frac{3}{2} \left( \frac{1}{2} \dot{u} 
+ \frac{\dot{a}}{a}  \right) L^{-1}  \e^{-u/2} \Gamma^{\hat{0}} \psi \eta
+ \tilde{f} \chi \psi \eta \nonumber \\
=&  L^{-1}  \e^{-u/2} \left\{  \Gamma^{\hat{0}}  \partial_0 \psi + \frac{1}{ a \left( t \right)} 
\Gamma^{\hat{i}} \partial_i \psi \right\} \eta 
+ \left\{  \partial_5 \eta + u^{\prime} \left( t,w \right) \eta   \right\} \Gamma^{\hat{5}} \psi 
+ \tilde{f} \chi \left( w \right) \eta \psi \nonumber \\
& + L^{-1}  \e^{-u/2} \left\{  \partial_0 \eta + \frac{3}{2} \left(  \frac{1}{2} \dot{u} 
+ \frac{\dot{a}}{a} \right) \eta \right\} \Gamma^{\hat{0}} \psi \, .
\end{align}
The explicit form of the spin connections are given in (\ref{B7}) in Appendix \ref{Ap2}. 
Let $\tilde{C}_{1} \left( w \right)$ and $D_{1} \left( t \right) $ be arbitrary functions. 
Then by decomposing $\eta \left( t ,w \right)$ as 
\begin{align}
\label{YT3}
\eta \left( t, w \right) & \overset{\mathrm{def}}{=} \zeta \left( t \right) \lambda \left( w \right) g \left( t, w \right)  \\
g \left( t, w \right)     & \overset{\mathrm{def}}{=} \exp \left[  \tilde{C}_1 \left( w \right) + D_1 \left( t \right) - u \left( t, w \right) \right] \, ,
\end{align}
and by assuming $\Gamma^{\hat{5}} \psi = \pm \psi$, we find 
\begin{align}
\label{YT4}
\Gamma^{M} \nabla_{M} \Psi + \tilde{f} \chi \left( w \right) \Psi =&  L^{-1}  \e^{-u/2} 
\left\{  \Gamma^{\hat{0}}  \partial_0 \psi + \frac{1}{ a \left( t \right)} \Gamma^{\hat{i}} 
\partial_i \psi \right\} \eta \nonumber \\
& \pm \zeta g \left\{  \partial_5 \lambda \left( w \right)  
+ \tilde{C}^{\prime}_{1} \left( w \right) \lambda \left( w \right) 
\pm \tilde{f} \chi \left( w \right) \lambda \left( w \right) \right\} \psi \nonumber \\
& + L^{-1}  \e^{-u/2} \lambda g \left\{  \partial_0 \zeta \left( t \right) 
+ \left(  \frac{3}{2} \frac{\dot{a}}{a} - \frac{1}{4} \dot{u} \left( t, w \right) 
+ \dot{D}_{1} \left( t \right)  \right) \zeta \left( t \right)  \right\} \Gamma^{\hat{0}} \psi \, .
\end{align}
By expressing $\e^{u\left(  t, w \right)}$ as 
$\e^{u\left(  t, w \right)} = T \left( t \right) W \left( w \right)$ and by using the Dirac equation, 
we find 
\begin{align}
\label{YT5}
\lambda \left( w \right) &= C_3 \exp \left[  - \tilde{C}_1 \left( w \right) \mp \tilde{f}  \int \mbox{d}w \, \chi \left( w \right) \right] \\
\zeta \left( t \right)  &= C_4 a^{-3/2} T^{1/4} \exp \left[ - D_1 \left( t \right)  \right] \, .
\end{align}
Thus we obtain  
\begin{align}
\label{YT6}
\eta \left( t, w  \right) =& C_3 C_4 a^{-3/2} T^{1/4} \exp \left[ - u \left( t, w \right) \mp 
\tilde{f}  \int \mbox{d} w \,   \chi \left( w \right) \right] \nonumber \\
=& C_5 a^{-3/2} T^{-3/4} W^{-1} \exp \left[  \mp \tilde{f} \int \mbox{d} w \,  \chi \left( w \right)  \right] \, .
\end{align}
Because $\chi \left( w \right) = w$, we find the explicit form of $\eta \left( t, w  \right)$, 
\begin{equation}
\label{YT7}
\eta \left( t, w  \right) = C_5 a^{-3/2} T^{-3/4} W^{-1} \exp \left[ \mp \frac{\tilde{f}}{2} w^2 \right] \, .
\end{equation}
Therefore the condition of localization is given by 
\be
\label{YT8}
I  = \int^{+ \infty }_{- \infty } \mbox{d} w \,  \e^{3u/2 }\left| \eta  \right|^2 
= C_{5}^2 a^{-3} \int^{+ \infty}_{- \infty} \mbox{d} w \, W^{-1/2} \exp \left[  \mp \tilde{f} w^2 \right] < \infty \, .
\ee

We now consider some examples given in \cite{Toyozato:2012zh, Higuchi:2014bya}. 
\begin{itemize} 
\item {\bf Example 1.} 
$W \left( w \right) = \mbox{ e} ^{-w^2/w^2_0}$, $T \left(  t \right) = T_1 t^{1-3h_0} + T_2 t^{-2h_0}$.
This example corresponds to the FRW universe filled with the perfect fluid whose Equation of state parameter $w$ is constant. 
Then we find 
\begin{align}
\label{YT9}
I = C^2_5 a^{-3} \int^{+ \infty}_{-\infty} \mbox{d} w \,  \exp \left[ \left( \frac{1}{2 w^2_0} \mp \tilde{f} \right) w^2 \right]\, ,
\end{align}
which tells the chirality of the fermions localized on the domain wall, 
\begin{equation}
\label{YT10}
\begin{cases}
\tilde{f}  &> \displaystyle \frac{1}{2 w^2_0}  \quad \quad \quad      \mbox{for}  \quad \quad \Gamma^{\hat{0}} \psi = \psi \\
\tilde{f}  &< \, -  \displaystyle  \frac{1}{2 w^2_0} \quad \quad     \mbox{for}  \quad \quad \Gamma^{\hat{0}} \psi = - \psi
\end{cases}\, .
\end{equation}
\item {\bf Example 2} $u \left( t, w \right) = -3 H_0 t + \log\left\{  - \alpha \left( w \right) + \exp \left[ H_0 t \right] \right\} + \beta \left( w \right)$. 
This example  describes the de Sitter universe embedded in five dimensions. 
We now decompose $u$ as $u \left( t, w \right) = T \left( t \right) W \left( w \right)$ if and only if $\alpha \left( w \right) = \alpha_0 = \mbox{const}$. In that case, we obtain 
\begin{align}
I \propto \int^{+ \infty}_{- \infty} \mbox{d} w \, \exp \left[  -\frac{1}{2} \beta \left( w \right) \mp \tilde{f} w^2 \right] \, .
\end{align}
Now for simplicity, let $\alpha_0 = 0$. In this case, we have shown in previous work 
\cite{Toyozato:2012zh} that the ghost-less condition is 
$\beta^{\prime \prime} \left( w \right) < 0$. 
For instance, the case $\beta \left( w \right) = - \beta^2_0 w^2 $ satisfies this condition. 
In this case, the condition of localization is given by 
\begin{equation}
\begin{cases}
\tilde{f}  &> \displaystyle \frac{\beta^2_0}{2}  \quad \quad \quad      \mbox{for}  \quad \quad \Gamma^{\hat{0}} \psi = \psi \\
\tilde{f}  &< \, -  \displaystyle  \frac{\beta^2_0}{2} \quad \quad     \mbox{for}  \quad \quad \Gamma^{\hat{0}} \psi = - \psi
\end{cases} \, .
\end{equation}
\end{itemize}
Thus we have shown that the fermion can be localized on the domain wall and the localized 
fermion can be chiral or anti-chiral. 
As we have mentioned, the attempt in \cite{George:2008vu} to localize the fermion on the domain wall has been failed. 
This is because in \cite{George:2008vu}, it was assumed that the warp factor does not depend 
on the time but we have used the time-dependent warp factor and we have succeeded to 
show the localization of the chiral fermion.

\section{Localization of Vector Field \label{IV}}

We now consider the localization of the vector field whose action is given by 
\be
\label{V1}
S_V = \int d^5 x \sqrt{-g} \left\{ - \frac{1}{4} F_{MN} F^{MN} - \frac{1}{2} m(\chi)^2 A_M A^M \right\}\, , \quad 
F_{MN} = \partial_M A_N - \partial_N A_M\, .
\ee
In the background (\ref{metricR}) with (\ref{FRWmetric0}), we choose $k=0$ and $\e^{u\left(  t, w \right)} = T \left( t \right) W \left( w \right)$ as in (\ref{YT5}). 
Then the background metric has the following form: 
\be
\label{metricRB}
ds^2 = dw^2 + L^2 W(w) T(t) ds_\mathrm{FRW}^2 \, , \quad 
ds_\mathrm{FRW}^2 = - dt^2 + a(t)^2 \sum_{i=1}^3 \left( dx^i \right)^2 \, .
\ee
Then the action (\ref{V1}) reduces to 
\begin{align}
\label{V2}
S_V =& \int d^5 x \left\{ \frac{1}{2} L^2 W(w) T(t) a(t)^3 F_{50}^2 
 - \frac{1}{2} L^2 W(w) T(t) a(t) F_{5i}^2 
+ \frac{1}{2}a(t) F_{0i}^2 - \frac{1}{4} a(t)^{-1} F_{ij}^2 \right. \nn
& \left. - \frac{1}{2} m(\chi)^2 \left( L^4 W(w)^2 T(t)^2 a(t)^3 A_5^2 
 - L^2 W(w) T(t) a(t)^3 A_0^2 + L^2 W(w) T(t) a(t) A_i^2 \right) \right\}\, .
\end{align}
By the variation of $A_5$, $A_0$, and $A_i$, we obtain the following equations,
\begin{align}
\label{V3}
0 = & L^2 W(w) \partial_0 \left( T(t) a(t)^3 \left( \partial_5 A_0 - \partial_0 A_5 \right) \right) - L^2 W(w) T(t) a(t) \left(\partial_5 \partial_i A_i - \partial_i^2 A_5 \right) \nn
& - m(\chi)^2 L^4 W(w)^2 T(t)^2 a(t)^3 A_5 \, , \\
\label{V4}
0 = & - L^2 T(t) a(t)^3 \partial_5 \left( W(w) \left( \partial_5 A_0 - \partial_0 A_5 \right) \right) + a(t) \left( \partial_0 \partial_i A_i - \partial_i^2 A_0 \right) + m(\chi)^2 L^2 W(w) T(t) a(t)^3 A_0\, , \\
\label{V5}
0 = & L^2 T(t) a(t) \partial_w \left( W(w) \left( \partial_5 A_i - \partial_i A_5 \right) \right) - \partial_0 \left( a \left( \partial_0 A_i - \partial_i A_0 \right) \right) - a(t)^{-1}  \left( \partial_i \partial_j A_j - \partial_j^2 A_i \right) \nn
& - m(\chi)^2 L^2 W(w) T(t) a(t) A_i \, .
\end{align}
Then by assuming 
\be
\label{V6} 
A_5 =0 \, , \quad 
A_\mu = X(w) C_\mu \left( x^\nu \right)\, , \quad 
\mu,\nu=0,1,2,3\, ,
\ee
and by choosing 
\be
\label{V7}
m\left( \chi=w \right)^2 = \frac{\left( W (w) X'(w) \right)'}{W(w) X(w)}\, ,
\ee
Eqs.~(\ref{V3}), (\ref{V4}), and (\ref{V5}) have the following forms, respectively. 
\begin{align}
\label{V8}
0 =& \partial_5 X(w) \left\{ \partial_0 \left(T(t) a(t)^3 C_0 \right) - T(t) a(t) \partial_i C_i \right\}\, , \\
\label{V9}
0 = & \partial_0 \partial_i C_i - \partial_i^2 C_0 \, , \\
\label{V10}
0 = &
\partial_0 \left( a(t) \left( \partial_0 C_i - \partial_i C_0 \right) \right) 
+ a(t)^{-1} \left( \partial_i \partial_j C_j - \partial_j^2 C_i \right) \, .
\end{align}
Eqs.~(\ref{V9}) and (\ref{V10}) are nothing but the field equations of the vector field in four dimensions. 
Eq.~(\ref{V8}) can be regarded as a gauge condition, which is a generalization of the Landau gauge, $\partial^\mu A_\mu = 0$.  

Then if we choose $X(w)$ decreases rapidly enough for large $\left|w \right|$, $A_\mu$ can be normalizable. Then by choosing $m(\chi)$ as in (\ref{V7}), the vector field can localize on the domain wall. 
In case $m(\chi)\neq 0$, we cannot consider the non-abelian gauge theory but if 
\be
\label{V11}
\left( W (w) X'(w) \right)' = 0\, ,
\ee
we obtain $m(\chi)=0$ and we might be able to consider the non-abelian gauge theory. 
In case that $W \left( w \right) = \mbox{ e} ^{-w^2/w^2_0}$, however, the vector field is not normalizable.

\section{Discussions \label{V}}

In this paper, 
based on the previous works \cite{Toyozato:2012zh, Higuchi:2014bya}, we have investigated the 
localization of the fields on the dynamical domain wall, where the four dimensional FRW universe 
is realized on the domain wall in the five dimensional space-time. 
Especially we have shown that the chiral spinor can localize on the domain wall, which has not 
been shown in the past works. 
An attempt in \cite{George:2008vu} to localize the fermion on the dynamical domain wall has 
been failed because the warp factor does not depend on the time. 
As we have seen, we have shown that by introducing the time-dependent warp factor, we can 
obtain the chiral fermion localized on the domain wall. 

The localized fields corresponding to the zero modes in the corresponding fields. There, of course, exist non-localized or non-confining modes as solutions of the field equations. When we consider the interactions between the fields, there might be mixing with these non-localized modes. Especially we have assumed that the background is fixed but in general, there could be some mixing with the modes from gravity and the scalar fields $\phi$ and $\chi$. 
%%%%%%%%%%%%%%%
Therefore although we have found solutions localized on the domain wall, all the solutions are not always localized.   
%%%%%%%%%%%%%%%

In (\ref{YT1})  and in (\ref{V1}), we have considered the Yukawa coupling and interaction only with $\chi$ and we have not included another scalar field $\phi$. 
This is because one of the purposes in this paper is to show the existence of the models where the localization can occur and we have considered simple models as possible. 
We may include another scalar field $\phi$ in the interactions but we can show that there could be models which generate the localization even if we include $\phi$.

\section*{Acknowledgments.}

This work is supported (in part) by 
MEXT KAKENHI Grant-in-Aid for Scientific Research on Innovative Areas ``Cosmic
Acceleration''  (No. 15H05890) and the JSPS Grant-in-Aid for Scientific Research (C) \# 23540296 (S.N.).

\appendix

\section{Explicit expressions of connections and curvatures in five dimensions  \label{Ap1}}

In this Appendix, we give explicit expressions for the connections and curvatures in five 
dimensional space-time, whose 
metric is given by
\begin{align}
\label{A1}
 g_{AB}=\begin{pmatrix}
 	-L^2 \e^{u(w,t)}&&&&\\
	 & L^2 \e^{u(w,t)}\frac{a(t)^2}{1-kr^2} & & & \\
	 & & L^2 \e^{u(w,t)}a(t)^2r^2 & & \\
	 & & & L^2 \e^{u(w,t)}a(t)^2r^2\sin^2\theta & \\
	 & & & & 1 
	\end{pmatrix}\, .
\end{align}\\
The connections are given by
\begin{align}
\label{A2}
&\Gamma^t_{tt}=\frac{1}{2}\dot{u}\, ,\quad
\Gamma^w_{tt}=\frac{1}{2}L^2\e^uu^\prime\, , \quad
\Gamma^r_{rt}=\Gamma^\theta_{\theta t}=\Gamma^\phi_{\phi t}=\frac{\dot{a}}{a}+\frac{1}{2}\dot{u} \, , \quad
\Gamma^t_{tw}=\Gamma^r_{rw}=\Gamma^\theta_{\theta w}=\Gamma^\phi_{\phi w}=\frac{1}{2}u^\prime\, , \nn
&\Gamma^t_{ij}=L^{-2} \e^{-u}\left(\frac{\dot{a}}{a}+\frac{1}{2}\dot{u}\right)g_{ij}\, , \quad
\Gamma^r_{rr}=\frac{kr}{1-kr^2}\, , \quad
\Gamma^w_{ij}=-\frac{1}{2}u^\prime g_{ij}\, , \quad
\Gamma^\theta_{\theta r}=\Gamma^\phi_{\phi r}=\frac{1}{r}\, , \nn
&\Gamma^r_{\theta\theta}=-r(1-kr^2)\, , \quad
\Gamma^\phi_{\phi\theta}=\cot\theta\, , \quad
\Gamma^r_{\phi\phi}=-r(1-kr^2)\sin^2\theta\, ,\quad
\Gamma^\theta_{\phi\phi}=-\cos\theta\sin\theta \, .
\end{align}
The Ricci curvatures have the following forms:
\begin{align}
R_{tt}&=\left[-\frac{1}{2}u^{\prime\prime}-u^{\prime2}+\frac{3}{2}L^{-2} \e^{-u}\left(\ddot{u}+\frac{\dot{a}\dot{u}}{a}
+2\frac{\ddot{a}}{a}\right)\right]g_{tt}\, ,\nn
R_{ij}&=\left[-\frac{1}{2}u^{\prime\prime}-u^{\prime2}+\frac{1}{2}L^{-2} \e^{-u}\left(\ddot{u}+5\frac{\dot{a}\dot{u}}{a}
+2\frac{\ddot{a}}{a}+4\frac{\dot{a}^2}{a^2}+\dot{u}^2+4\frac{k}{a^2}\right)\right]g_{ij} \, ,\nn
R_{ww}&=-2u^{\prime\prime}-u^{\prime2} \, \nn
R_{tw}&=-\frac{3}{2}\dot{u}^\prime \, .
\end{align}
The scalar curvatures is
\begin{align}
\label{A3}
 R=-4u^{\prime\prime}-5u^{\prime2}+3L^{-2} \e^{-u}\left(\ddot{u}+\frac{1}{2}\dot{u}^2
+3\frac{\dot{a}\dot{u}}{a}+2\frac{\ddot{a}}{a}+2\frac{\dot{a}^2}{a^2}+2\frac{k}{a^2}\right)\, .
\end{align}

\section{Explicit form of spin connection \label{Ap2}}

The explicit forms of the spin connections in five dimensions are given in this Appendix. 

By using the metric in (\ref{metricR}), we define the vierbein field $e^{\hat{B}}$, 
\begin{equation}
\label{B1}
g_{MN}  
\overset{\mathrm{def}}{=} e^{\hat{A}}_{\, \, \, M} \eta_{\hat{A} \hat{B}} e^{\hat{B}}_{\, \, \, N}
\end{equation}
The components of the vierbein field $e^{\hat{B}}$ is given by
\begin{equation}
\label{B2}
e^{\hat{A}}_{\, \, \, M} =
\begin{bmatrix}
L  \e^{u/2} & 0 & 0 & 0 & 0\\
0 & L  \e^{u/2} a & 0 & 0 & 0\\
0 & 0 & L  \e^{u/2} a & 0 & 0\\
0 & 0 & 0 & L  \e^{u/2} a & 0\\
0 & 0 & 0 & 0 & 1
\end{bmatrix}\, .
\end{equation}
We also find 
\begin{equation}
\label{B3}
e_{\hat{A} M} \overset{\mathrm{def}}{=} \eta_{\hat{A} \hat{B}} \,  e^{\hat{B}}_{\, \, \, M} =
\begin{bmatrix}
 -L  \e^{u/2} & 0 & 0 & 0 & 0\\
0 & L  \e^{u/2} a & 0 & 0 & 0\\
0 & 0 & L  \e^{u/2} a & 0 & 0\\
0 & 0 & 0 & L  \e^{u/2} a & 0\\
0 & 0 & 0 & 0 & 1
\end{bmatrix} \, ,
\end{equation}
and 
\begin{equation}
\label{B4}
e_{\hat{A} }^{\, \, \, \, \,M} \overset{\mathrm{def}}{=}  e_{\hat{A} N} g^{MN} =
\begin{bmatrix}
L^{-1}  \e^{-u/2} & 0 & 0 & 0 & 0\\
0 & L^{-1}  \e^{-u/2} a^{-1} & 0 & 0 & 0\\
0 & 0 & L^{-1}  \e^{-u/2} a^{-1} & 0 & 0\\
0 & 0 & 0 & L^{-1}  \e^{-u/2} a^{-1} & 0\\
0 & 0 & 0 & 0 & 1
\end{bmatrix}\, .
\end{equation}
Then we define the Ricci rotation coefficients and spin connections as follows, 
\begin{align}
\label{B5}
\Omega_{ABM} & \overset{\mathrm{def}}{=} \left(  \partial_A  e^{\hat{C}}_{\, \, \, B} - \partial_B  e^{\hat{C}}_{\, \, \, A}  \right)  e_{\hat{C} M} = - \Omega_{BAM} \, , \\
\label{B6}
\omega_{ABM} & \overset{\mathrm{def}}{=} \displaystyle - \frac{1}{2} \left(  \Omega_{ABM}  -\Omega_{BMA} -\Omega_{MAB}  \right) = - \omega_{BAM} \, .
\end{align}
The non-vanishing components are given by 
\begin{equation}
\label{B7}
\begin{cases}
\Omega_{050} &= - \Omega_{500} = \displaystyle \frac{1}{2} u^{\prime} L^2  \e^{u} \\
\Omega_{0ij} &= - \Omega_{i0j} = \,  \displaystyle \left(  \frac{1}{2} \dot{u} a^2 + \dot{a} a \right) L^2  \e^{u} \delta_{ij} \, , \nn
\Omega_{5ij} &= - \Omega_{i5j} = \displaystyle \frac{1}{2} u^{\prime} L^2  \e^{u} a^2 \delta_{ij} \, ,
\end{cases}
\end{equation}
and
\begin{equation}
\label{B8}
\begin{cases}
\omega_{050} &= - \omega_{500} = - \displaystyle \frac{1}{2} u^{\prime} L^2  \e^{u} \, , \nn
\omega_{0ij} &= - \omega_{i0j} = \,  - \displaystyle \left(  \frac{1}{2} \dot{u} a^2 + \dot{a} a \right) L^2  \e^{u} \delta_{ij} \, , \nn
\omega_{5ij} &= - \omega_{i5j} = - \displaystyle \frac{1}{2} u^{\prime} L^2  \e^{u} a^2 \delta_{ij} \, .
\end{cases}
\end{equation}

\end{document}